\documentclass[sigconf]{acmart}

\usepackage{array}
\usepackage{tabularx}
\usepackage{subcaption}
\usepackage{graphicx}
\usepackage[most]{tcolorbox}
\usepackage[utf8]{inputenc}
\usepackage{textgreek}
\usepackage{ulem}

\usepackage{booktabs}
\usepackage{siunitx}
\usepackage{pgfplots}
\usepackage{pgfplotstable}
\pgfplotsset{compat=1.18}

\usepackage{subcaption}

\usepgfplotslibrary{statistics}

\tcbset{
  rqbox/.style={
    colback=gray!10!white,   
    colframe=black!70,       
    fonttitle=\bfseries,
    coltitle=black,
    boxrule=0.8pt,
    arc=3mm,
    left=4mm,
    right=4mm,
    top=2mm,
    bottom=2mm,
    before skip=10pt,
    after skip=10pt
  }
}

\tcbset{
  rqbox/.style={
    colback=gray!10!white,   
    colframe=pink!70,       
    fonttitle=\bfseries,
    coltitle=black,
    boxrule=0.8pt,
    arc=3mm,
    left=4mm,
    right=4mm,
    top=2mm,
    bottom=2mm,
    before skip=10pt,
    after skip=10pt
  }
}
\usepackage{xcolor}
\usepackage{ifthen}
\usepackage{soul}

\usepackage{pgf-pie}

\newif\ifpienumberinlegend
\pgfkeys{/number in legend/.code=
    \expandafter\let\expandafter\ifpienumberinlegend
    \csname if#1\endcsname
    \ifpienumberinlegend

    \def\beforenumber##1\afternumber{}%
    \fi,
    /number in legend/.default=true
}

\newboolean{showcomments}
\setboolean{showcomments}{true} 

\ifthenelse{\boolean{showcomments}}
{
  \newcommand{\nbc}[3]{
    \colorbox{#3}{\bfseries\sffamily\scriptsize\textcolor{white}{#1}}
    {\textcolor{#3}{\sf$\blacktriangleright$\textit{#2}$\blacktriangleleft$}}
  }
}
{
  \newcommand{\nbc}[3]{}
}

\copyrightyear{2026}
\acmYear{2026}
\setcopyright{cc}
\setcctype{by}
\acmConference[MSR '26]{23rd International Conference on Mining Software Repositories}{April 13--14, 2026}{Rio de Janeiro, Brazil}
\acmBooktitle{23rd International Conference on Mining Software Repositories (MSR '26), April 13--14, 2026, Rio de Janeiro, Brazil}
\acmPrice{}
\acmDOI{10.1145/3793302.3793611}
\acmISBN{979-8-4007-2474-9/2026/04}

\begin{document}

\title{Why Are AI Agent–Involved Pull Requests (Fix-Related) Remain Unmerged? An Empirical Study}

\author{Khairul Alam}
\affiliation{%
  \institution{University of Saskatchewan, Canada}
  \city{}
  \state{}
  \country{}
}
\email{kha060@usask.ca}

\author{Saikat Mondal}
\affiliation{%
  \institution{University of Saskatchewan, Canada}
  \city{}
  \state{}
  \country{}
}
\email{saikat.mondal@usask.ca}

\author{Banani Roy}
\affiliation{%
  \institution{University of Saskatchewan, Canada}
  \city{}
  \state{}
  \country{}
}
\email{banani.roy@usask.ca}

\renewcommand{\shortauthors}{Trovato et al.}

\begin{abstract}

Autonomous coding agents (e.g., OpenAI Codex, Devin, GitHub Copilot) are increasingly used to generate fix-related pull requests (PRs) in real-world software repositories. However, their practical effectiveness depends on whether project maintainers accept and merge these contributions. In this paper, we present an empirical study of AI agent–involved fix-related PRs, examining both their integration outcomes, latency, and the factors that hinder successful merging. We first analyze $8,106$ fix-related PRs authored by five widely used AI coding agents from the AIDEV-POP dataset to quantify the proportions of PRs that are merged, closed without merging, or remain open. We then conduct a manual analysis of a statistically significant sample of 326 closed but unmerged PRs, spending approximately 100 person-hours to construct a structured catalog of 12 failure reasons. Our results indicate that test case failures and prior resolution of the same issues by other PRs are the most common causes of non-integration, whereas build or deployment failures are comparatively rare. Overall, our findings expose key limitations of current AI coding agents in real-world settings and highlight directions for their further improvement and for more effective human-AI collaboration in software maintenance.
\end{abstract}


\maketitle
\section{Introduction}

Modern autonomous coding agents, such as OpenAI Codex, Devin, Cursor, and GitHub Copilot, are increasingly capable of performing end-to-end software development tasks, including feature implementation, debugging, bug fixing, documentation, and even code review, with minimal human oversight \cite{bouzenia2024repairagent, li2025rise, wang2025agents}. These agents are already integrated into thousands of real-world software development projects and often function as virtual teammates that assist developers throughout the software development lifecycle \cite{li2025rise}. Their ability to generate PRs, respond to review feedback, and iteratively refine code highlights the growing potential of automated agents to support software engineering workflow \cite{hassan2024towards}.

Despite their rapid adoption, the effectiveness of such agents ultimately depends on whether their contributions are \textit{accepted and integrated} into real software systems \cite{zakharov2025ai}. In modern development workflows, PRs are the primary mechanism for reviewing, negotiating, and merging changes, making PR acceptance a key indicator of successful human-AI collaboration \cite{watanabe2025use}. Among different contribution types, \textit{fix-related PRs} are particularly critical, as they directly impact software reliability and user trust, and are therefore subject to stricter correctness, testing, and validation requirements.

Prior studies on automated program repair \cite{zhang2024systematic, bouzenia2024repairagent} and LLM-based bug fixing \cite{tian2024evaluating} demonstrate that, although AI agents can often generate syntactically correct patches, many fixes fail to address root causes, introduce unintended side effects, or lack sufficient test validation \cite{liu2024marscode, grundy2005deployed}. As a result, such fixes frequently fail to pass code review and are ultimately not merged \cite{monperrus2014critical, sobania2023analysis, jimenez2023swe}. 
For example, in pull request \texttt{\#553} to the Azure/CloudShell\footnote{\url{https://github.com/Azure/CloudShell/pull/553}} repository, an AI agent proposed a fix to resolve a version mismatch among Microsoft.Graph PowerShell modules by aligning dependency versions so that modules can be imported without assembly errors. Despite multiple automated commits in response to reviewer feedback, the PR repeatedly failed CI tests, did not satisfy reviewer expectations, and was ultimately closed without merging by a maintainer. This case illustrates that generating a plausible patch alone is often insufficient for successful integration when tests or project constraints remain unaddressed. Such observations motivate a systematic investigation into how AI agent-authored fix-related PRs perform in real-world review workflows and why many of them are unmerged.

Existing research on AI-assisted software development primarily emphasizes benchmark-based evaluations and task-level success rates \cite{jimenez2023swe, sobania2023analysis}, or developer productivity outcomes \cite{peng2023impact, dohmke2023economic}, often abstracting away the downstream PR review and integration process. As a result, there is limited empirical understanding of how AI agent-generated fixes perform within real-world PR workflows under human review and project constraints \cite{bacchelli2013expectations, rigby2012contemporary}, and, in particular, why fix-related PRs authored by AI agents are closed without being merged. Such closed but unmerged PRs represent wasted computational effort and increased review overhead for maintainers, while simultaneously exposing practical limitations in agent reasoning, validation strategies, and interaction with human reviewers.

To address this gap, we conduct an empirical study of fix-related PRs authored by five widely used AI coding agents using the AIDEV-POP dataset \cite{li_2025_16919272}. We analyze integration outcomes at scale and construct a structured catalog of failure reasons through in-depth manual analysis of closed but unmerged PRs. In particular, we investigate the following research questions:

\begin{itemize}
    \item[\textbf{RQ1}]\textbf{\textit{(Outcome and Latency Analysis):} What proportion of AI agent–involved fix-related PRs are merged, closed without merging, or remain open? What is the submission-to-merge time for merged PRs?}
    Autonomous coding agents increasingly submit PRs to real-world repositories, but their effectiveness depends on whether these contributions are successfully integrated. Merge outcomes provide a concrete signal of integration, reflecting correctness, test compliance, and alignment with project conventions, while submission-to-merge time captures additional integration effort such as review and coordination overhead. However, there is limited empirical evidence on how often AI agent–authored fix-related PRs are merged, closed without merging, or remain open, or how long successful merges take. Accordingly, this RQ evaluates agent effectiveness by analyzing both integration outcomes and submission-to-merge time of fix-related PRs in real-world workflows.
    
    \item[\textbf{RQ2}]\textbf{\textit{(Failure Analysis):} Why are AI agent–involved fix-related PRs closed without merging?}
    PRs that are closed without merging represent concrete failure points in human-AI collaboration, where AI agent-generated contributions do not meet project or maintainer expectations. Analyzing such unmerged PRs enables the construction of a structured catalog of failure causes that explain why AI-generated fixes fail to integrate in practice. Understanding these failure causes is essential for diagnosing the practical limitations of current AI coding agents and informing improvements to agent design, PR review, and integration workflows.
    
\end{itemize}

\noindent By jointly analyzing integration outcomes and failure causes of AI agent–authored fix-related PRs, our study clarifies where and why such contributions fail in practice, informing both agent improvement and more effective human-AI collaboration.

\smallskip
\noindent The \textbf{replication package} is available in our online appendix \cite{random_2025_18105922}.

\section{Methodology}
\label{sec:methodology}

We adopt a mixed-methods approach, combining quantitative analysis of PR integration outcomes with qualitative analysis of closed but unmerged PRs to construct a catalog of failure reasons.

\subsection{Dataset Selection}
\label{subsec:dataset}

We use the AIDEV-POP dataset \cite{li_2025_16919272} released in November 2025, a large-scale collection of AI-authored PRs mined from GitHub repositories. The dataset contains 33,596 PRs, of which 8,106 are classified as fix-related and authored by five widely used AI coding agents: OpenAI Codex (4,338), GitHub Copilot (1,993), Devin (1,249), Cursor (411), and Claude Code (115). We focus exclusively on fix-related PRs because they directly address defects or incorrect behavior and consequently undergo more rigorous review, testing, and validation processes. This makes them particularly suitable for evaluating the practical effectiveness of AI agents in real-world collaborative software development.

\subsection{Analysis of PR Integration Outcomes (RQ1)}
\label{subsec:methodology-rq1}

To assess the integration outcomes of AI agent-authored fix-related PRs, each PR is classified into one of three outcome states: \textit{merged}, \textit{closed without merging}, or \textit{open} based on its status at the time of data collection. We report both the \textit{overall} outcome distribution and the \textit{agent-wise} distributions to examine differences in integration outcomes across AI coding agents.

For merged PRs, we analyze \textit{time-to-merge} to characterize integration latency. Specifically, we compute time-to-merge as the difference (in hours) between the PR creation (\texttt{created\_at}) and closing timestamp (\texttt{closed\_at}) recorded by GitHub. We then analyze the distribution of time-to-merge and compare integration latency across different AI coding agents to examine agent-level differences in how quickly fix-related PRs are merged in practice.


\subsection{Analysis of Failure Causes in Closed but Unmerged PRs (RQ2)}

To address RQ2, we focus on fix-related PRs that were closed without merging. From the 2,113 unmerged cases identified in RQ1, we randomly sample 326 PRs, which provides a 95\% confidence level with a 5\% margin of error \cite{Wang-SOEdit-TSE2018, mondal2023automatic}. This sample size is sufficient for statistically representative qualitative analysis.

We then conduct a manual qualitative analysis using an open coding approach. Two authors independently label each PR, spending approximately \textit{100 person-hours} in total, following an initial calibration phase, in which they jointly analyzed 20 PRs to establish a shared understanding of the coding procedure and criteria. The calibration PRs are excluded from the final sample of 326 PRs. For each PR, the annotators examine the PR description, code changes, review comments (when available), commit history, and closure context to identify the primary reason for the unsuccessful merge outcome. Failure categories are derived inductively from the data rather than predefined.

To assess coding reliability, we compute Cohen’s kappa \cite{blackman2000interval}, obtaining a value of 0.82, indicating substantial inter-rater agreement. All disagreements are resolved through discussion to reach a consensus. Through this process, we construct a structured catalog of failure causes that explain why AI agent–authored fix-related PRs are closed without being merged.


\section{Study Findings}
\label{subsec:findings}

This section presents the empirical findings of our study.

\subsection{Findings of RQ1: Integration Outcomes of Fix-Related PRs}
\label{subsec:findings-rq1}

Table~\ref{tab:rq1-outcomes-p1} summarizes the integration outcomes of fix-related PRs authored by different AI coding agents. Overall, a majority of fix-related PRs are successfully merged (65.0\%), indicating that AI agents can frequently produce acceptable fixes in real-world projects. However, the outcomes vary substantially across agents. OpenAI Codex exhibits a notably high merge rate (81.6\%), whereas GitHub Copilot and Devin show much lower merge rates (42.4\% and 42.9\%, respectively) and a higher prevalence of closed-but-unmerged PRs. In particular, more than half of Devin's fix-related PRs are closed without merging, highlighting significant integration challenges. These differences suggest that while AI-generated fixes are often viable, their likelihood of successful integration depends strongly on the agent and on how its contributions align with project expectations and review workflows.

\begin{table}[!htb]
\centering
\caption{Integration outcomes of fix-related PRs authored by AI coding agents}
\label{tab:rq1-outcomes-p1}
\resizebox{3.2in}{!}{%
\begin{tabular}{lrrrr}
\toprule
\textbf{AI Agent} & \textbf{Total PRs} & \textbf{Merged} & \textbf{Closed w/o Merge} & \textbf{Open} \\
\midrule
OpenAI Codex   & 4,338 & 3,539 (81.6\%) & 616 (14.2\%) & 183 (4.2\%) \\
GitHub Copilot & 1,993 & 845 (42.4\%)   & 713 (35.8\%) & 435 (21.8\%) \\
Devin          & 1,249 & 536 (42.9\%)   & 675 (54.0\%) & 38 (3.0\%) \\
Cursor         & 411   & 281 (68.4\%)   & 84 (20.4\%)  & 46 (11.2\%) \\
Claude Code    & 115   & 66 (57.4\%)    & 25 (21.7\%)  & 24 (20.9\%) \\
\midrule
\textbf{Total} & 8,106 & 5,267 (65.0\%) & 2,113 (26.1\%) & 726 (8.9\%) \\
\bottomrule
\end{tabular}
}
\end{table}

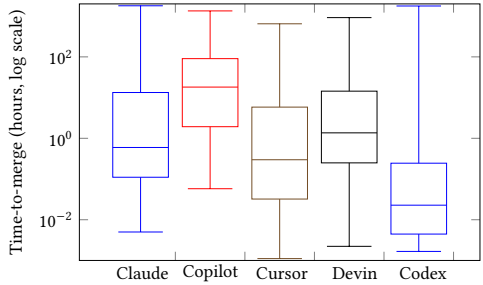
\begin{figure}[t]
\centering
\resizebox{2.5in}{!}{%
\begin{tikzpicture}
\begin{axis}[
  width=\linewidth,
  height=6cm,
  ymode=log,
  ymin=1e-3, ymax=2e3,
  ylabel={Time-to-merge (hours, log scale)},
  xtick={1.5,2.5,3.5,4.5,5.5},
  xticklabels={Claude, Copilot, Cursor, Devin, Codex},
  x tick label style={rotate=0, anchor=east, yshift=-6pt},
  grid style={line width=.1pt, draw=gray!30},
  major grid style={line width=.2pt, draw=gray!40},
  boxplot/draw direction=y,
  title style={font=\small}
]

\addplot+[
  boxplot prepared={
    lower whisker=0.005,
    lower quartile=0.1105555556,
    median=0.5952777778,
    upper quartile=13.32319444,
    upper whisker=1797.698611
  },
] coordinates {};

\addplot+[
  boxplot prepared={
    lower whisker=0.05777777778,
    lower quartile=1.928611111,
    median=18.11111111,
    upper quartile=90.50472222,
    upper whisker=1345.510278
  },
] coordinates {};

\addplot+[
  boxplot prepared={
    lower whisker=0.001111111111,
    lower quartile=0.03222222222,
    median=0.2975,
    upper quartile=5.835277778,
    upper whisker=649.1961111
  },
] coordinates {};

\addplot+[
  boxplot prepared={
    lower whisker=0.002222222222,
    lower quartile=0.2492361111,
    median=1.361111111,
    upper quartile=14.38673611,
    upper whisker=921.0388889
  },
] coordinates {};

\addplot+[
  boxplot prepared={
    lower whisker=0.001666666667,
    lower quartile=0.004444444444,
    median=0.02277777778,
    upper quartile=0.2447222222,
    upper whisker=1776.576944
  },
] coordinates {};

\end{axis}
\end{tikzpicture}%
}
\vspace{-0.6em}
\caption{Distribution of time-to-merge for merged fix-related PRs across AI coding agents (log scale).}
\label{fig:rq1-boxplot-merge-latency}
\end{figure}

We further analyze the time required for AI agent–authored fix-related PRs to be successfully merged. Fig. \ref{fig:rq1-boxplot-merge-latency} shows the distribution of merge latency across different AI agents. The results indicate substantial variability both within and across agents, with most merged PRs integrated quickly, while a smaller fraction experiences prolonged delays. In particular, OpenAI Codex exhibits consistently shorter merge times with a tightly concentrated interquartile range, whereas Copilot and Devin show broader distributions and heavier upper tails, reflecting greater variability in integration latency. Overall, the figure highlights that while many AI-generated fixes are merged rapidly, merge latency is strongly influenced by the agent and the complexity of the review process.

\subsection{Findings of RQ2: Failure Causes of Closed but Unmerged PRs}
\label{subsec:findings-rq2}

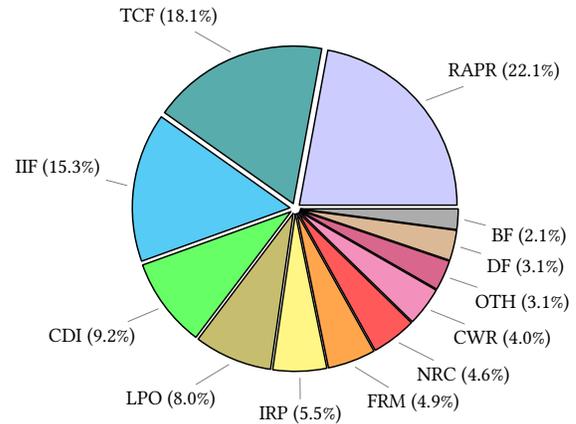
\begin{figure}[!htb]
	\centering
    	\resizebox{3in}{!}{
        \begin{tikzpicture}
        \pie[explode=0.1, text=pin, number in legend, sum = auto,    color={
          blue!20,
          teal!65,
          cyan!55,
          green!60,
          olive!55,
          yellow!60,
          orange!70,
          red!65,
          magenta!55,
          purple!60,
          brown!55,
          gray!65
        }]
            {
            22.1/\Large{RAPR (22.1\%)},
            18.1/\Large{TCF (18.1\%)},
            15.3/\Large{IIF (15.3\%)},
            9.2/\Large{CDI (9.2\%)},
            8.0/\Large{LPO (8.0\%)},
            5.5/\Large{IRP (5.5\%)},
            4.9/\Large{FRM (4.9\%)},
            4.6/\Large{NRC (4.6\%)},
            4.0/\Large{CWR (4.0\%)},
            3.1/\Large{OTH (3.1\%)},
            3.1/\Large{DF (3.1\%)},
            2.1/\Large{BF (2.1\%)}
            }
        \end{tikzpicture}
        \label{user-based-rollback-inconsistencies}
        }
    \vspace{-1em}
	\caption{Distribution of failure causes (R1–R12) for closed but unmerged fix-related PRs.}
	\label{fig:failure-causes-distribution}
	\vspace{-4mm}
\end{figure}

Our manual analysis of closed-but-unmerged fix-related PRs resulted in a structured catalog of \textbf{12 distinct failure reasons}. Figure~\ref{fig:failure-causes-distribution} illustrates the distribution of these reasons across the analyzed 326 PRs. Overall, the findings indicate that unsuccessful integration arises from a combination of \textit{technical shortcomings} and \textit{process-related factors}. We describe each failure reason below.

\noindent \textbf{(R1) Resolved by Another Pull Request (RAPR).}
This is the most frequent cause, accounting for 22.1\% (72 PRs). In these cases, the reported issue was resolved by a different PR that was merged earlier or in parallel. We identified such cases by tracing PR discussions and linked issues and verifying that the issue was closed due to another merged PR. For example, the PR \textit{``Fix JsonValue serialization InvalidCastException in Orleans JsonCodec''} was closed after the issue was resolved by a separate merged PR, \textit{``Improve support for serializing System.Text.Json types.''}

\noindent \textbf{(R2) Test Case Failures (TCF).}
18.1\% (59 PRs) were closed because the proposed changes failed existing tests or introduced new test failures. Reviewers explicitly mentioned failing tests as the primary reason for non-integration, such as in the PR \textit{``Fix debug expression hover for TypeScript non-null assertions (!).''}

\noindent \textbf{(R3) Incorrect or Incomplete Fixes (IIF).}
We find 15.3\% (50 PRs) in which the proposed fixes were incorrect, incomplete, or failed to resolve the reported issue. Reviewers indicated that the changes did not constitute a valid fix or introduced incorrect behavior. An example is \textit{``Fix division by zero handling inconsistency in arithmetic optimization''}, which was closed after reviewers determined the fix was invalid.

\noindent \textbf{(R4) Fix Rejected by Maintainers (FRM).}
In 4.9\% (16 PRs), fixes underwent review and revision but were ultimately rejected by maintainers. Although AI agents addressed the reviewer feedback, maintainers decided not to accept the final solutions. For instance, \textit{``Fix task restart to re-read tasks.json configuration''} was closed after multiple review iterations.

\noindent \textbf{(R5) Deployment Failures (DF).}
3.1\% (10 PRs) were closed due to deployment or runtime failures encountered during integration or release. Maintainers reported that the changes could not be deployed, as in \textit{``Fix smithery build config and bundle OpenAPI spec.''}

\noindent \textbf{(R6) Build Failures (BF).}
We identify 2.1\% (7 PRs) where proposed changes triggered build failures, typically reflected by failing CI checks. Maintainers explicitly cited build errors as the reason for closure, for example in \textit{``Fix fs.writeSync options parsing.''}

\noindent \textbf{(R7) Low Priority or Obsolete Issue (LPO).}
In 8.0\% (26 PRs), the reported issues were deprioritized or rendered obsolete by subsequent changes in the codebase, making the fixes unnecessary. For example, \textit{``Fix DataProduct assets persistence issue by removing premature sorting''} was closed as subsequent codebase changes resolved the issue.

\noindent \textbf{(R8) Incomplete Review Process (IRP).}
5.5\% (18 PRs) experienced partial review activity that did not reach a conclusion. Reviewers provided initial feedback, but the review process was not completed, and the PRs were eventually closed, such as in \textit{``Fix duplicate and unfitting slugs in resourceDefinition.json.''}

\noindent \textbf{(R9) Closed Due to Inactivity (CDI).}
We find 9.2\% (30 PRs) that were automatically closed due to prolonged inactivity. These PRs showed no further updates within the project-defined time window and were closed by repository automation.

\noindent \textbf{(R10) Closed Without Explicit Rationale (CWR).}
In 4.0\% (13 PRs), PRs were closed without any comments or explanation, leaving the closure rationale unclear. For example, \textit{``Fix module resolution for observability''} was closed unmerged without explanation.

\noindent \textbf{(R11) No Review Conducted (NRC).}
4.6\% (15 PRs) were closed without any reviewer engagement, even though AI agents explicitly requested review. These cases reflect a lack of review participation rather than technical evaluation.

\noindent \textbf{(R12) Other Reasons (OTH).}
The remaining 3.1\% (10 PRs) were closed for miscellaneous reasons, including repository archival, infrastructure or authentication issues, lack of substantive code changes, unexpected agent errors, or closure to submit a cleaner or revised PR.

\begin{table}[t]
\centering
\small
\caption{Agent-wise distribution of failure causes for closed-but-unmerged fix-related PRs.
Values are shown as count (percentage within each agent).}
\label{tab:agent-wise-failure-causes}
\resizebox{\linewidth}{!}{%
\begin{tabular}{lrrrrr}
\toprule
\textbf{Failure Cause} & \textbf{Claude} & \textbf{Copilot} & \textbf{Cursor} & \textbf{Devin} & \textbf{Codex} \\
\midrule
Resolved by Another PR & 5 (27.8\%) & 47 (31.8\%) & 5 (26.3\%) & 7 (14.0\%) & 8 (8.8\%) \\
Test Case Failures      & 0 (0.0\%)  & 6 (4.1\%)   & 0 (0.0\%)  & 3 (6.0\%)  & 50 (54.9\%) \\
Incorrect / Incomplete Fixes & 0 (0.0\%) & 30 (20.3\%) & 0 (0.0\%) & 4 (8.0\%) & 13 (14.3\%) \\
Closed Due to Inactivity & 1 (5.6\%) & 2 (1.4\%) & 0 (0.0\%) & 27 (54.0\%) & 0 (0.0\%) \\
Low Priority / Obsolete & 4 (22.2\%) & 8 (5.4\%) & 0 (0.0\%) & 0 (0.0\%) & 4 (4.4\%) \\
Incomplete Review      & 0 (0.0\%)  & 18 (12.2\%) & 0 (0.0\%) & 0 (0.0\%) & 4 (4.4\%) \\
Fix Rejected            & 0 (0.0\%)  & 13 (8.8\%) & 0 (0.0\%) & 4 (8.0\%) & 13 (14.3\%) \\
No Review Conducted     & 1 (5.6\%)  & 7 (4.7\%)  & 1 (5.3\%) & 2 (4.0\%) & 4 (4.4\%) \\
Closed w/o Rationale    & 3 (16.7\%) & 1 (0.7\%)  & 8 (42.1\%) & 1 (2.0\%) & 0 (0.0\%) \\
Deployment Failures      & 0 (0.0\%)  & 0 (0.0\%)  & 2 (10.5\%) & 3 (6.0\%) & 5 (5.5\%) \\
Build Failures          & 1 (5.6\%)  & 1 (0.7\%)  & 0 (0.0\%)  & 1 (2.0\%) & 4 (4.4\%) \\
Other Reasons          & 1 (5.6\%)  & 5 (3.4\%)  & 1 (5.3\%)  & 0 (0.0\%) & 3 (3.3\%) \\
\midrule
\textbf{Total PRs}             & \textbf{18} & \textbf{148} & \textbf{19} & \textbf{50} & \textbf{91} \\
\bottomrule
\end{tabular}
}
\vspace{-0.8em}
\end{table}

\smallskip
We further analyze agent-wise failure causes by normalizing each category by the total number of closed-but-unmerged PRs per agent (Table~\ref{tab:agent-wise-failure-causes}). For OpenAI Codex, failures are dominated by test case failures (50 out of 91, 54.9\%), indicating validation-related issues rather than review or prioritization. In contrast, Devin is primarily affected by inactivity-related closures (27 out of 50, 54.9\%), reflecting challenges sustaining engagement during longer review cycles. GitHub Copilot shows several notable failure profiles, with issues resolved by other PRs as the largest category (47 out of 148, 31.7\%), suggesting redundancy and timing effects alongside technical shortcomings. For Cursor and Claude Code, failures are relatively few and skew toward prioritization- or redundancy-related causes rather than systematic technical errors.

\smallskip
\noindent\textbf{Synthesis of Failure Causes.}
To better understand the nature of unsuccessful integration, we group the identified failure reasons into three high-level categories: \textit{technical}, \textit{process-related}, and \textit{priority-related} causes. \textbf{Technical causes} include cases where the proposed changes were incorrect or incomplete (IIF), failed test cases (TCF), or triggered build or deployment failures (BF, DF). These failures indicate limitations in the agents’ ability to generate fully correct and validation-ready fixes. \textbf{Process-related causes} capture breakdowns in the pull-request workflow, including fixes rejected after review (FRM), incomplete review processes (IRP), absence of review (NRC), closures without explicit rationale (CWR), and closures due to inactivity (CDI). Such cases highlight challenges in review interaction, feedback handling, and sustained engagement. Finally, \textbf{priority-related causes} include issues resolved by another PR (RAPR) or deemed low priority or obsolete (LPO), reflecting redundancy, timing mismatches, or shifting maintainer priorities rather than intrinsic technical flaws. Together, this synthesis shows that while many failures stem from technical shortcomings, a substantial portion arises from workflow and prioritization dynamics inherent to real-world collaborative software development.

\section{Key Findings}


\noindent \textbf{Integration reflects encouraging outcomes with notable limitations.}
Across 8,106 fix-related PRs, 65.0\% are merged, while 26.1\% are closed without merging and 8.9\% remain open, indicating that a substantial fraction of AI agent–authored fixes remain unmerged.

\noindent \textbf{Integration success varies notably across agents.}
Merge rates differ considerably by agent, ranging from 81.6\% (OpenAI Codex) to 42.9\% (Devin), suggesting that real-world effectiveness may depend on agent design and alignment with project workflows.

\noindent \textbf{Validation-related issues are more prevalent than build failures.}
Among closed-but-unmerged PRs, test case failures (18.1\%) and incorrect or incomplete fixes (15.3\%) occur substantially more often than build (2.1\%) or deployment failures (3.1\%), pointing to limitations in validation readiness rather than basic compilability.

\noindent \textbf{Workflow and timing factors play a non-trivial role.}
A notable share of unmerged PRs are closed because the issue was resolved by another PR (22.1\%) or due to inactivity (9.2\%), indicating that process-related dynamics and timing effects can affect integration outcomes alongside technical quality.

\noindent \textbf{Merge latency varies substantially across agents.}
For merged PRs, submission-to-merge times show considerable variability both within and across agents, with some agents exhibiting shorter and more concentrated merge times than others, suggesting differences in review effort and coordination overhead during integration.

\section{Conclusion and Future Work}

This paper presents an empirical study of AI agent-involved fix-related PRs using the AIDEV-POP dataset. Analyzing 8,106 PRs authored by five widely used AI coding agents, we find that while many AI-generated fixes are merged, a substantial fraction are closed without merging or remain open. Integration success varies across agents, and non-merge outcomes are frequently associated with validation-related issues (e.g., failing tests or incomplete fixes) as well as workflow- and timing-related factors. A qualitative analysis of closed-but-unmerged PRs further yields a structured catalog of 12 failure causes, suggesting that syntactically plausible fixes alone are often insufficient for successful integration.
As future work, we plan to investigate test- and context-aware agent designs that better support automated validation, review interactions, and alignment with project-specific expectations, and to examine maintainer–agent interactions over extended review cycles.

\section*{Acknowledgement}
This research is supported in part by the Natural Sciences and Engineering Research Council of Canada (NSERC), and by the industry-stream NSERC CREATE in Software Analytics Research (SOAR)
\bibliographystyle{ACM-Reference-Format}
\bibliography{sample-base}
\end{document}